# Perturbation Theory for the Systems of Ordinary Linear Differential Equations with Periodical Coefficients

A.G. Kvirikadze, M.D. Zviadadze, T.V. Tavdgiridze, I.G. Tavelidze

E. Andronikashvili Institute of Physics, Academy of Sciences of Georgia

Tamarashvili 6, Tbilisi, 0177 Georgia

m.zviadadze@mail.ru

The method, proposed in the given work, allows the application of well developed standard methods used in quantum mechanics for approximate solution of the systems of ordinary linear differential equations with periodical coefficients.

1. Let us consider the system of n ordinary linear differential equations of the first order in time:

$$\frac{dy_j(t)}{dt} = \sum_{j'=1}^{n} a_{jj'}(t) y_{j'}(t) + f_j(t), \quad j = 1,2,\ldots n \qquad (1)$$

where $y_j(t), f_j(t)$ and $a_{jj'}(t)$ are the complex functions of actual variable **t**. Introducing the vector functions-columns $\vec{y}(t), \vec{f}(t)$ with $y_j(t), f_j(t)$ components and matrix $\hat{a}(t)$ with matrix elements $a_{jj'}(t)$, one can write the system (1) in vector form:

$$\frac{d\vec{y}(t)}{dt} = \hat{a}(t)\vec{y}(t) + \vec{f}(t) \qquad (2)$$

It is well-known [1], that the solution of equation (2) has the form:

$$\vec{y}(t) = \hat{U}(t)\vec{y}_0 + \hat{U}(t)\int_0^t dt' \hat{U}^{-1}(t')\vec{f}(t'), \qquad (3)$$

where the term $\hat{U}(t)\vec{y}_0$ is the solution of homogeneous equation

$$\frac{d\vec{y}(t)}{dt} = \hat{a}(t)\vec{y}(t) \qquad (4)$$

with initial condition $\vec{y}(0) \equiv \vec{y}_0$, and the fundamental matrix $\hat{U}(t)$ satisfies equation

$$\frac{d\hat{U}(t)}{dt} = \hat{a}(t)\hat{U}(t), \quad \hat{U}(0) = I, \qquad (5)$$

$I$ is the unit matrix.

In the case, when $a_{jj'}(t)$ coefficients are the periodical functions with $T = 2\pi/\omega$ period, i.e.

$$a_{jj'}(t+T) = a_{jj'}(t), \qquad (6)$$

according to Flock theory [1], matrix $\hat{U}(t)$ can be presented in the form:

$$\hat{U}(t) = \tilde{U}(t)e^{\hat{\mu}t}, \qquad (7)$$

where

$$\tilde{U}(t+T) = \tilde{U}(t), \quad \tilde{U}(0) = I, \qquad (8)$$

and matrix $\hat{\mu}$ is not time-dependent.

When the diagonalization of $\hat{\mu}$ is possible,[1] the Flock theory, finally, comes to the statement, that homogeneous equation (4) has n of linearly independent solutions of the following form:

$$\vec{y}^{(\ell)}(t) = \vec{\psi}^{(\ell)}(t)e^{-\mu_\ell t}, \quad \vec{\psi}^{(\ell)}(t+T) = \vec{\psi}^{(\ell)}(t), \quad \ell = 1,2,...,n. \qquad (9)$$

It is easy to prove, that the matrix, constructed using $\psi^{(\ell)}(t)$ and $\mu_\ell$

$$\hat{U}(t) = \begin{pmatrix} \psi_1^{(1)}, & \psi_1^{(2)}, & ..., & \psi_1^{(n)} \\ \psi_2^{(1)}, & \psi_2^{(2)}, & ..., & \psi_2^{(n)} \\ ..., & ..., & ..., & ... \\ \psi_n^{(1)}, & \psi_n^{(2)}, & ..., & \psi_n^{(n)} \end{pmatrix} \cdot \begin{pmatrix} e^{-\mu_1 t}, & 0, & ..., & 0 \\ 0, & e^{-\mu_2 t}, & ..., & 0 \\ ..., & ..., & ..., & ... \\ 0, & ..., & 0 & e^{-\mu_n t} \end{pmatrix} \cdot \hat{C} \qquad (10)$$

is the fundamental matrix of Eq. (2), if matrix $\hat{C}$ is determined from $\hat{U}(0) = \hat{C}$ condition.

Solution (9) of Eq. (4) reduce to the solution of infinite system of linear homogeneous algebraic equations for Fourier-components of unknown $\psi_j^{(\ell)}(t)$ functions, and the estimation of $\mu_\ell$ values – to the estimation of eigenvalues of corresponding infinite matrix, which, in the general case, is practically impossible. For this reason, it is especially important to find efficient methods of approximate solution of equations (2) and (4).

There are a lot of methods of approximate solution of Eq. (4) [1, 2, 4]. All of them are mainly applicable for the case, when $\hat{a}(t)$ can be presented as

$$\hat{a}(t) = \hat{a}_0(t) - \hat{V}(t) \qquad (11)$$

Here, one can think, that matrix $\hat{V}(t)$ is, in a sense, small as compared to $a_0(t)$ and consider it as perturbation (minus sign is chosen for the sake of convenience). However, it should be noted, that all these methods have the difficulties, connected with appearance and exclusion of secular terms and with understanding of the structure of a series of perturbation theories in general.

The proposed method is free of the mentioned shortcomings and allows to analyze relatively easily a series of perturbation theories in general.

---

[1] If matrix $\hat{\mu}$ is not diagonalized, it can be presented to Jordan form [3]. This case requires a separate consideration.

2. Substitution of (11) into (4) gives:

$$\frac{d\vec{y}(t)}{dt} = [\hat{a}_0(t) - \hat{V}(t)]\vec{y}(t). \qquad (12)$$

Let us introduce operators

$$\hat{H} = \frac{d}{dt} - \hat{a}(t) = \hat{H}_0 + \hat{V}, \quad \hat{H}_0 = \frac{d}{dt} - \hat{a}_0(t), \quad \frac{d}{dt} \equiv I\frac{d}{dt}. \qquad (13)$$

It is easy to show, that if we know n of linearly independent periodical eigenvectors $\vec{\psi}^{(j)}(t)$ of $\hat{H}$ operator:

$$\hat{H}\vec{\psi}^{(j)}(t) = \mu_j \vec{\psi}^{(j)}(t), \quad \vec{\psi}^{(j)}(t+T) = \vec{\psi}^{(j)}(t), \quad j = 1,2,...,n \qquad (14)$$

and the corresponding eigenvalues $\mu_j$, the matrix (10), constructed by using them, is a fundamental matrix of Equation (12).

For approximate solution of Eq. (14) it is necessary to know the full system of periodical eigenvectors $\vec{\phi}_{jk}(t)$ and eigenvalues $\aleph_{jk}$ of $\hat{H}_0$ operator

$$\hat{H}_0 \vec{\phi}_{jk}(t) = \aleph_{jk} \vec{\phi}_{jk}(t), \quad j = 1,2,...,n; \quad k = 0,\pm 1, \pm 2 \pm,...,\pm\infty. \qquad (15)$$

Values $\vec{\phi}_{jk}(t)$ and $\aleph_{jk}$ are easy to find by means of n of linearly independent solutions of equation

$$\frac{d\vec{y}_0(t)}{dt} = \hat{a}_0(t)\vec{y}_0(t), \qquad (16)$$

having the form:

$$\vec{y}_0^{(j)}(t) = \vec{\phi}^{(j)}(t)e^{-\aleph_j t}, \quad \vec{\phi}^{(j)}(t+T) = \phi^{(j)}(t), \quad j = 1,2,...,n \qquad (17)$$

in correspondence with Flock theory. It is easy to check, that the vectors

$$\vec{\phi}_{jk}(t) = e^{ik\omega t}\vec{\phi}^{(j)}(t) \qquad (18)$$

form the full system of periodical eigenvectors $\hat{H}_0$ with eigenvalues

$$\aleph_{jk} = \aleph_j + ik\omega, \qquad (19)$$

where $i$ is the imaginary unit, and $\omega = 2\pi/T$

3. Vectors $\vec{\phi}_{jk}(t)$ form the basis in linear space $E$ of periodical **n**-dimensional complex vector functions. The space $E*$ of linear functionals $\vec{\phi}^{(+)}$ on $E$, conjugated to $E$, can be also realized in the form of periodical vector functions $\vec{\phi}^{(+)}(t)$, determining the value of $\vec{\phi}^{(+)}$ functional on vector $\vec{\chi} \in E$ by the rule:

$$<\varphi^{(+)}|\vec{\chi}>=\frac{1}{T}\int_{-T/2}^{T/2}dt(\vec{\varphi}^{(+)}(t),\vec{\chi}(t))=\frac{1}{T}\int_{-T/2}^{T/2}dt\sum_{j=1}^{n}\varphi_j^{(+)}(t)\chi_j(t), \qquad (20)$$

where $(\vec{\varphi}^{(+)}(t),\vec{\chi}(t))$ is the ordinary scalar product of **n**-dimensional vectors.

In conjugated $E^*$ space one can introduce $\vec{\varphi}_{jk}^{(+)}$ basis, biorthogonal, to $\vec{\varphi}_{jk}$ basis:

$$<\vec{\varphi}_{jk}^{(+)}|\vec{\varphi}_{j'k'}>=\delta_{jj'}\delta_{kk'}, \qquad (21)$$

where $\delta_{jj'},\delta_{kk'}$ are the Kroneker symbols. Further, let us use the Dirac symbols [5], considering the vectors of $E$ space as ket-vectors, and the vectors of conjugated $E^*$ space - as bra-vectors. In particular, the basis $\vec{\varphi}_{jk}(t)$ vectors we will write in the form of ket-vectors $|jk>$, and the basis $\vec{\varphi}_{jk}^{(+)}(t)$ vectors of conjugated $E^*$ space – in the form of bra-vectors $<jk|$.

For the future, it is necessary to consider operators $\widehat{A}$ acting in $E$ and having the form:

$$\widehat{A}=\begin{pmatrix} \hat{A}_{11}, & \hat{A}_{12}, & ..., & \hat{A}_{1n} \\ \hat{A}_{21}, & \hat{A}_{22}, & ..., & \hat{A}_{2n} \\ ..., & ..., & ..., & ..., \\ \hat{A}_{n1}, & \hat{A}_{n2}, & ..., & \hat{A}_{nn} \end{pmatrix}, \qquad (22)$$

where the matrix elements $\hat{A}_{jj'}$ are the operators themselves, acting on the function of $t$:

$$\hat{A}\cdot\vec{\chi}(t)=\begin{pmatrix} \sum_j \hat{A}_{1j}\chi_j(t) \\ \sum_j \hat{A}_{2j}\chi_j(t) \\ .......... \\ \sum_j \hat{A}_{nj}\chi_j(t) \end{pmatrix}. \qquad (23)$$

Matrix elements of $\hat{A}$ operator in the basis $|j'k'>,<jk|$ are:

$$A_{jk,j'k'}=<jk|\hat{A}|j'k'>=\frac{1}{T}\int_{-T/2}^{T/2}dt\cdot(\vec{\varphi}_{jk}^{(+)}(t),\hat{A}\vec{\varphi}_{j'k'}(t))=\frac{1}{T}\int_{-T/2}^{T/2}dt\sum_{\alpha,\beta=1}^{n}\varphi_{jk}^{(+)\alpha}(t)\hat{A}_{\alpha\beta}\varphi_{j'k'}^{(\beta)}(t)$$

(24)

4. Solutions of Eq.(14) are directly connected with the study of the properties of resolvent operator $\hat{G}(\mu)$ (or Green function) of $\widehat{H}$ operator, determined by expression:

$$\hat{G}(\mu)=(\mu-\hat{H})^{-1}, \quad \hat{H}=\hat{H}_0+\hat{V}. \qquad (25)$$

Together with $\hat{G}(\mu)$ let us consider operators

$$\hat{F}(\mu)=\hat{G}(\mu)\cdot\frac{1}{\hat{g}(\mu)}, \quad \hat{R}(\mu)=\hat{V}\hat{F}(\mu), \quad \hat{g}(\mu)=\sum_{j=1}^{n}\sum_{k=-\infty}^{\infty}G_{jk}(\mu)|jk><jk|, \qquad (26)$$

where

$$G_{jk}(\mu) = < jk | (\mu - \hat{H})^{-1} | jk > \tag{27}$$

In [5, 6] it is shown that by means of $\hat{F}(\mu)$ and $\hat{R}(\mu)$ operators one can find eigenfunctions and eigenvalues of $\hat{H}$ operator. Indeed, by using identity $(\mu - \hat{H})\hat{G}(\mu) = \hat{I}$ and Eq. (26), we will obtain

$$(\mu - \hat{H})\hat{F}(\mu) | jk > = \frac{1}{\hat{g}(\mu)} \cdot | jk > = \frac{1}{G_{jk}(\mu)} \cdot | jk > . \tag{28}$$

Here $\hat{I}$ is the unit operator in $E$ space. Taking into account the condition of completeness of the system of basis vectors

$$\sum_{j=1}^{n} \sum_{k=-\infty}^{\infty} | jk >< jk | = \hat{I},$$

for diagonal matrix elements of $\hat{R}(\mu)$ operator we find

$$R_{jk}(\mu) = < jk | (\hat{H} - \hat{H}_0 | jk > = < jk | (\mu - \hat{H}_0)\hat{F}(\mu) | jk > - < jk | (\mu - \hat{H})\hat{F}(\mu) | jk > =$$

$$= (-\aleph_{jk}) < jk | \hat{F}(\mu) | jk > - \frac{1}{G_{jk}(\mu)},$$

i.e.

$$\mu = \aleph_{jk} + R_{jk}(\mu) + \frac{1}{G_{jk}(\mu)}, \tag{29}$$

because $< jk | \hat{F}(\mu) | jk > = 1$. From the determination of $G_{jk}(\mu)$ it is seen, that as $\mu \to \mu_{jk}$, where $\mu_{jk}$ is the eigenvalue of $\hat{H}$ operator, $G_{jk}(\mu) \to \infty$. At $\mu = \mu_{jk}$, Eqs. (28) and (29) take the form

$$\hat{H}\hat{F}(\mu_{jk}) | jk > = \mu_{jk} \hat{F}(\mu_{jk}) | jk >, \tag{30}$$

$$\mu_{jk} = \aleph_{jk} + R_{jk}(\mu_{jk}). \tag{31}$$

As it follows from (30), $\hat{F}(\mu_{jk}) | jk >$ is the eigenfunction of $\hat{H}$ operator with eigenvalue $\mu_{jk}$, determined by Eq. (31). Thus, the shift operator $\hat{R}(\mu)$ determines the change of eigenvalue $\aleph_{jk}$ of $\hat{H}_0$ operator under the action of perturbation $\hat{V}$.

In order to obtain the equation for $\hat{F}(\mu)$ operator in the form suitable for iteration, we use the operator equality

$$\hat{A}^{-1} = \hat{B}^{-1} + \hat{B}^{-1}(\hat{B} - \hat{A})\hat{A}^{-1}$$

valid for any operators $\hat{A}$ and $\hat{B}$, for which there are inverse operators $\hat{A}^{-1}$ and $\hat{B}^{-1}$. By choosing $\hat{A} = \hat{g}(\mu)(\mu - H)$ and $\hat{B} = \hat{g}(\mu)(\mu - \hat{H}_0 - \hat{O}(\mu))$, we obtain the following equation for operator $\hat{F}(\mu)$:

$$\hat{F}(\mu) = \frac{1}{\mu - \hat{H}_0 - \hat{O}(\mu)} \cdot \frac{1}{\hat{g}(\mu)} + \frac{1}{\mu - \hat{H}_0 - \hat{O}(\mu)}(\hat{V} - \hat{O}(\mu))\hat{F}(\mu). \qquad (32)$$

5. Equation (32) contains arbitrary operator $\hat{O}(\mu)$, the choice of which is dictated for reasons of convenience in each specific problem. Different choice of $\hat{O}(\mu)$ leads to different series of perturbation theory. The most familiar two choices lead to series of perturbation theory in the form of Vigner – Brilluene and in the form of Rayleigh-Schrodinger [6, 7]. Assuming:

$$\hat{O}(\mu) = R_{jk}(\mu) \cdot |jk><jk|, \qquad (33)$$

we obtain the expansion of $\vec{\psi}_{jk}$ eigenfunction of $\hat{H}$ operator in Vigner-Brilluene form. And indeed, acting by $\hat{F}(\mu)$ operator on $|jk>$ vector and using (32) and (29), we obtain:

$$\hat{F}(\mu)|jk> = \frac{1}{\mu - \hat{H}_0 - \hat{O}(\mu)}(\mu - \aleph_{jk} - R_{jk}(\mu))|jk> + \frac{1}{\mu - \hat{H}_0 - \hat{O}(\mu)}(\hat{V} - \hat{O}(\mu))\hat{F}(\mu)|jk>$$

(34)

where

$$\frac{1}{\mu - \hat{H}_0 - \hat{O}(\mu)} = \sum_{j=1}^{n}\sum_{k'=-\infty}^{\infty} \frac{1}{\mu - \aleph_{j'k'} - O_{j'k'}(\mu)}|j'k'><j'k'|,$$

$$O_{j'k'}(\mu) = R_{jk}(\mu)\delta_{j'j}\delta_{k'k}. \qquad (35)$$

Solution of Eq. (34) by means of iteration and consideration of (35) lead to the expansion of function (34) (at $\mu = \mu_{jk}$ it becomes eigenfunction of $\hat{H}$ operator) in series of perturbation theory in Vigner – Brilluene form:

$$\hat{F}(\mu)|jk> = |jk> + \sum_{j'k' \neq jk}\frac{1}{\mu - \aleph_{j'k'}}|j'k'><j'k'|\hat{V}|jk> +$$

$$+ \sum_{\substack{j'k' \neq jk \\ j''k'' \neq jk}}\frac{1}{\mu - \aleph_{j'k'}} \cdot \frac{1}{\mu - \aleph_{j''k''}}|j'k'><j'k'|\hat{V}|j''k''><j''k''|\hat{V}|jk> + \cdots \quad (36)$$

and to the equation for determination of $\mu_{jk}$ eigenvalue

$$\mu_{jk} = \aleph_{jk} + <jk|\hat{V}|j'k'> + \sum_{j'k'\neq jk} \frac{1}{\mu_{jk} - \aleph_{j'k'}} <jk|\hat{V}|j'k'><j'k'|\hat{V}|jk> + \cdots \quad (37)$$

In similar way, we can show, that the choice of

$$\hat{O}(\mu) = R_{jk}(\mu)|jk><jk| + (\mu - \aleph_{jk})\sum_{j'k'\neq jk}|j'k'><jk| \quad (38)$$

leads to the expansion of $\hat{F}(\mu)|jk>$ and $\mu_{jk}$ in series of perturbation theory in Rayleigh-Schrodinger form:[2]

$$\hat{F}(\mu)|jk> = |jk> + \sum_{j'k'\neq jk} \frac{1}{\aleph_{jk} - \aleph_{j'k'}} |j'k'><j'k'|\hat{V}|jk> + \sum_{\substack{j'k'\neq jk \\ j''k''\neq jk}} \frac{1}{\aleph_{jk} - \aleph_{j'k'}} \cdot \frac{1}{\aleph_{jk} - \aleph_{j''k''}} \cdot$$

$$\cdot |j'k'>(<j'k'|\hat{V}|j''k''> - \delta_{j'j}\delta_{k'k''}<jk|\hat{V}|jk>)<j''k''|\hat{V}|jk> + \cdots \quad (39)$$

$$\mu_{jk} = \aleph_{jk} + <jk|\hat{V}|jk> + \sum_{j'k'\neq jk} \frac{1}{\aleph_{jk} - \aleph_{j'k'}} <jk|\hat{V}|j'k'><j'k'|\hat{V}|jk>. \quad (40)$$

It is important to note that for exact application of the methods used in quantum mechanics, in the given approach it is not necessary to have Hilbert space and self-adjoined operator $\hat{H}$, required in quantum mechanics and only the existence of $|jk>$ basis is enough. The solutions obtained by expressions (36), (37) or (39), (40) for **n** of latent $\vec{\psi}^{(j)}$ vectors and for corresponding eigenvalues $\mu_j$ of $\hat{H}$ operator:

$$\vec{\psi}^{(j)} = \hat{F}(\mu_j)|j0>, \quad \mu_j = \mu_{j0}, \quad j=1, 2, ..., n; \quad k=0 \quad (41)$$

determine the fundamental matrix (10) and give the possibility to solve non-homogeneous equation (2) by Eq. (3).

As it is seen from Eqs. (36) and (39), the proposed method of approximate solution of Eqs. (2) or (4) allows to avoid the difficulties, connected with appearance and exclusion of secular terms and makes it possible to change the series of perturbation theory, to single out the main terms, to make partial summing using the corresponding graphical technique, etc.

---

[2] Perturbation theory in Vigner – Brilluene form is useful in the case, when there is degeneration in the spectrum of $\hat{H}_0$ operator or when the difference in certain eigenvalues is small, while for application of Rayleigh-Schrodinger theory it is necessary to make additional re-determination of the functions of zero approximation.